\documentclass[aps,prl,twocolumn,superscriptaddress,shownopacs,nobalancelastpage]{revtex4}
\usepackage{amsmath,amssymb,latexsym}
\usepackage{graphicx}
\graphicspath{{../fig/}}
\usepackage{txfonts,mathrsfs}
\usepackage{comment}

\DeclareSymbolFont{symbols}{OMS}{cmsy}{m}{n}
\DeclareSymbolFont{largesymbols}{OMX}{cmex}{m}{n}
\newcommand{\bm}[1]{\boldsymbol #1}

\newcommand{\kk}{{\bm k}}
\newcommand{\KK}{{\bm K}}

\newcommand{\QQ}{{\bm Q}}

\begin{document}

\title{Ultra-fast photo-carrier relaxation in Mott insulators with short-range spin correlations}

\author{Martin~Eckstein}
\affiliation{Max Planck Research Department for Structural Dynamics, University of Hamburg-CFEL, 22761 Hamburg, Germany}
\email{martin.eckstein@mpsd.cfel.de}
\author{Philipp~Werner}
\affiliation{Department of Physics, University of Fribourg, 1700 Fribourg, Switzerland}

\begin{abstract}
We compute the time-resolved photoemission spectrum after photo-doping in a two-dimensional Mott-Hubbard insulator. We find that the relaxation rate of high-energy photo-doped electrons in the paramagnetic phase scales with the strength of the 
nearest-neighbor spin correlations, which implies a pronounced increase of the relaxation times with temperature and excitation density. Finite doping, in contrast, opens additional scattering channels and leads to a faster relaxation. To obtain our results we have implemented a nonequilibrium version of the dynamical cluster approximation (DCA), which, in contrast to single-site dynamical mean-field theory, captures the effect of 
short-range correlations.
\end{abstract}

\date{\today}

\maketitle

Photo-doping, i.e., the production of mobile carriers in
an insulator with a short laser pulse, allows 
to trigger an ultra-fast insulator-to-metal transition. In Mott insulators, the strong interactions between electrons, spins and lattice can cause highly non-trivial effects in the relaxation processes that lead to the transient metallic state and its decay \cite{Ogasawara00, Iwai03, Okamoto07, Okamoto2010, Okamoto2011, Dean2011, Mitrano2014}. For example, in the cuprate Nd$_2$CO$_4$ Okamoto {\em et al.}~\cite{Okamoto2010} found a rapid relaxation of the Drude peak in the optical conductivity within less than $40$fs, followed by the formation of a mid-gap absorption band which indicates the existence of polaronic quasiparticles. The photo-excited metallic state in Mott insulators persists up to several picoseconds, which is long on electronic timescales, but orders of magnitude faster than in semiconductors. 
An understanding of the ultra-fast relaxation processes in which photo-doped carriers loose their initial kinetic energy is of interest for several reasons: 
On a fundamental level,  the dominant scattering channel may reveal information about the interactions which are mediated by the corresponding fluctuations (spins or phonons) and provide a different view on the complex phase diagram of correlated materials \cite{Gianetti11,DalConte2012,Wentao2013}. Moreover, dissipation resulting from the scattering of mobile carriers with other degrees of freedom is a limiting factor for transport at large fields \cite{Amaricci2012, Vidmar2011},
and similarly for photo-carrier diffusion and separation in multi-layered structures with large internal field gradients \cite{Eckstein2014Layer}. 

In Mott insulators with strong antiferromagnetic fluctuations and a relatively large gap, scattering with spins is a natural candidate for fast energy relaxation processes. (If the Mott gap is small, impact ionization is a competing process, in which a high energy carrier relaxes by exciting further carriers across the gap \cite{WernerAuger2014}.) A single carrier moving in an antiferromagnetically ordered background 
flips 
a spin in every hopping process. In equilibrium (and for more than one dimension) this gives rise to the binding of holes into string-states, evident in narrow spin polaron peaks in the spectral function \cite{Trugman1988, Strack1992, Sangiovanni2006, Bonca2007, Taranto2012}. In nonequilibrium, the same interaction causes a relaxation of carriers in an antiferromagnetic background on  a femtosecond timescale, which has been studied with exact diagonalization techniques \cite{Golez2014, Bonca2012, Mierzejewski2011, Kogoj2014}, the self-consistent Born approximation \cite{Iyoda2014}, and nonequilibrium dynamical mean-field theory \cite{Werner2012afm, Eckstein2014Layer}. 
Spin-charge interactions at low energies can lead to the formation of Mott excitons, which are relevant for the picosecond decay of the photo-excited state \cite{Lenarcic2013,Lenarcic2014}. 
While many numerical approaches work for a static antiferromagnetic background, a corresponding analysis of the paramagnetic phase, where spin-correlations are short-ranged and short-lived, is a challenge. This regime is especially important for the study of doped systems, which have no long-range magnetic order even at low temperatures. Doping reduces antiferromagnetic correlations, but at the same time the presence of low-energy mobile carriers opens additional scattering channels, so that the combined effect on the relaxation is not clear.

A powerful theoretical approach to study the dynamics of correlated systems is the nonequilibrium dynamical mean-field theory (DMFT) \cite{Freericks2006,REVIEW}. While DMFT has been used extensively to study the antiferromagnetic phase with long-range order  \cite{Tsuji2012,Werner2012afm,Eckstein2014Layer,Mentink2014a,Mentink2014b}, it neglects short-range spin-correlations in the paramagnetic phase, and thus is mainly applicable to lattices with high coordination number \cite{Metzner1989}, as well as systems at high temperature and high fluence in which spin-correlations have been melted away (see below). Relaxation processes in this case are restricted to electron-electron scattering, leading to a very slow or absent redistribution of spectral weight \cite{Eckstein11,Eckstein2012c,Moritz2012}. For the current analysis we thus have to go beyond DMFT. We use a nonequilibrium version of the dynamical-cluster approximation (DCA) \cite{Maier2005}, which maps a lattice problem to a small cluster embedded in a self-consistent bath and thus captures short-range correlations in the lattice. 

{\bf Model --} 
To study the interaction of photo-doped carriers with spin-fluctuations we consider the Hubbard model on a two-dimensional square lattice
\begin{equation}
\label{Hubbard}
H=-\!\!\sum_{\langle ij \rangle,\sigma}t_{ij}c^\dagger_{i\sigma}c_{j\sigma}
+
U\sum_i
\big(
n_{i\uparrow}-\tfrac12
\big)
\big(
n_{i\downarrow}-\tfrac12
\big)
+
\mu\sum_{i\sigma} n_{i\sigma}.
\end{equation}
Here the operators $c_{i\sigma}$ create an electron with spin $\sigma$ at lattice site $i$, $U$ is the on-site interaction, and $t_{ij}$ is the nearest neighbor hopping.  The laser excitation is modeled by a few-cycle electric field pulse of the form $E(t)=E_0e^{-4.6 (t-t_0)^2/t_0^2}\sin(\Omega t)$ with a frequency $\Omega$ comparable to the gap, polarized along the body-diagonal of the lattice ($t_0=2\pi/\Omega$). The electric field $\bm E(t)$ of the pump-laser pulse is incorporated into the Hamiltonian \eqref{Hubbard} by a Peierls phase, i.e., we choose $t_{ij}=t_{*}\exp(e \bm A(t)\bm r_{ij}/\hbar)$, where $\bm r_{ij}$ is the vector pointing from site $j$ to $i$, and $\bm A(t)$ is the vector potential from which the electric field is obtained by $\bm E(t)=- \partial_t \bm A(t)$. We measure energy in units of $t_{*}$, time in units of $1/t_*$ ($\hbar=1$), and the field in units of $t_{*}/ea$, where $a$ is the lattice spacing. Taking typical parameters for a cuprate Mott insulator ($U=3$eV, $t_{*}=0.25$eV, i.e., $U/$bandwidth$=1.5$), the unit of time is $2.6$fs, and temperature is measured in units of $t_{*}/k_{B}=2900$K. A quantitative modeling of both photo-doped holes and electrons in cuprates would require not only a next nearest neighbor hopping term (which is straightforward to incorporate), but also a multi-band model if particles are excited across the charge transfer gap. To study the qualitative effects of carrier relaxation in the presence of short-range spin fluctuations, we stick to the particle-hole symmetric model \eqref{Hubbard}.

The model is solved using the nonequilibrium formulation of DCA.  As for nonequilibrium DMFT \cite{Freericks2006,REVIEW}, the generalization of DCA to the time-domain can be done on a formal level by extending the theory from a Matsubara formalism to a Keldysh formalism \cite{TsujiDCA}, and we restrict the discussion to issues relevant for the implementation of the approach in the strong-coupling regime. The Brillouin zone of the lattice is divided into $L_c$ patches $P_\KK$, which in our case will be the patches around the high-symmetry points $\KK\in \{(0,0),(0,\pi),(\pi,0),(\pi,\pi)\}$. The lattice self-energy is approximated by a course-grained function $\Sigma_\kk(t,t') = \sum_{\KK} \Theta_{\KK,\kk} \Sigma_{\KK}(t,t')$ which is piecewise constant, i.e., $\Theta_{\KK,\kk} = 1$ if $\kk\in P_\KK$ and $0$ otherwise. The functions $\Sigma_\KK$ are obtained from an effective cluster model that consists of $2\times2=L_c$ sites embedded in a self-consistent bath. Formulated in momentum space, the action on the Keldysh contour is given by  
\cite{footnote02} 
\begin{align}
\label{DCAaction}
\mathcal{S}
&=
-i
\int_\mathcal{C} dt \,H_{c}(t)
-i
\int_\mathcal{C} dt dt'
\sum_{\KK\sigma}
c_{\KK\sigma}^\dagger \Delta_{\KK}(t,t')c_{\KK\sigma},
\\
H_{c} &= 
\frac{U}{L_c} \sum_{\KK,\KK',\QQ}\!\!
c_{\KK\uparrow}^\dagger
c_{\KK+\QQ\uparrow}
c_{\KK'\downarrow}^\dagger
c_{\KK'-\QQ\downarrow}
+
\sum_{\KK\sigma}
\bar \epsilon_\KK(t) c_{\KK\sigma}^\dagger
c_{\KK\sigma},
\end{align}
where $\Delta_\KK$ is the hybridization with the bath, and $\bar\epsilon_\KK(t)=\frac{L_c}{L}\sum_{\kk \in P_\KK} \epsilon_\kk(t)$ is the patch-averaged dispersion (which is time-dependent in the presence of electric fields). From the cluster model we compute the patch Green's functions  $G_{\KK}(t,t') = -i \langle T_\mathcal{C} c_{\KK}(t) c^\dagger_{\KK}(t')\rangle$ and the patch self-energies 
$\Sigma_{\KK}(t,t')$ which are related by the Dyson equation $G_{\KK}^{-1}(t,t') = i\partial_t +\mu -\bar\epsilon_{\KK}(t)-\Delta_{\KK}(t,t') - \Sigma_{\KK}(t,t')$. Momentum-dependent lattice Green's functions are computed with the course-grained self-energy $G_{\kk}^{-1}(t,t') = i\partial_t +\mu - \epsilon_{\kk}(t) - \Sigma_{\kk}(t,t')$, and the self-consistency is closed by the condition $G_\KK=\frac{L_c}{L}\sum_{\kk\in P_\KK} G_\kk$.

We use the lowest order self-consistent hybridization expansion (non-crossing approximation, NCA) to compute the Green's function $G_{\KK}$ from 
the 
action \eqref{DCAaction}. NCA is a diagrammatic expansion around the atomic limit ($\Delta_\KK=0$), which can be formulated in terms of pseudo-particles representing the many-body eigenstates $|m\rangle$ of the isolated cluster. The diagrammatic rules for NCA have been described in detail in Ref.~\cite{Eckstein2010nca}.  For a general multi-orbital case, the pseudo-particle Green's functions $\mathcal{G}$ and self-energies are matrices with flavor indices $m$, so that the solution of real-time Dyson equations for these matrix-valued two-time Green's functions can become numerically challenging.  (The effort scales like $(N_td)^3$, where $N_t$ is the number of elementary time-steps, and $d$ is the flavor-matrix dimension.) Taking into account quantum numbers $|m\rangle\equiv|N_{\uparrow},N_{\downarrow},\bm K;\alpha\rangle$ (particle numbers $N_{\uparrow,\downarrow}$, total momentum $\bm K$), the pseudo-particle Green's function's are block diagonal, $\mathcal{G}_{(N_{\uparrow},N_{\downarrow},\bm K;\alpha),(N_{\uparrow}',N_{\downarrow}',\bm K';\alpha')} \propto \delta_{N_\uparrow N'_\uparrow }  \delta_{N_\downarrow N'_\downarrow } \delta_{\bm K \bm K'}$, and the largest matrix dimension $d_\text{max}$ is $12$ ($N_\uparrow=N_\downarrow=2$, $\bm K=0$). The exponential scaling of $d_\text{max}$ with $L_c$ currently restricts the implementation of real-time NCA to the $2\times2$ cluster used here, which should be sufficient as a minimal model to capture short-range spin-fluctuations in an antiferromagnet. For the numerical solution we use a hybrid parallelized implementation, where Green's functions for different quantum numbers are distributed over several compute nodes, while the solution of the pseudo-particle Dyson equations 
uses a shared memory parallelization. 
\begin{figure}[tbp]
\centerline{%
\includegraphics[width=\columnwidth]{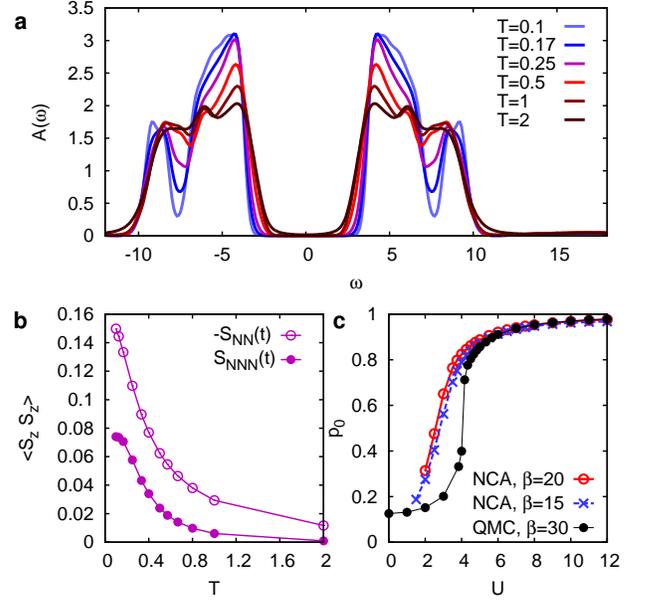}}
\caption{
a) Temperature dependent spectral function in the half-filled insulator for $U=12$.
b) Nearest neighbor (NN) and next-nearest neighbor (NNN) spin-correlations in the impurity model.
c) Occupation probability $p_0$ of the plaquette-singlet state (see text); QMC data are taken from Ref.~\cite{Gull2008}.
}
\label{fig1}
\end{figure}

In equilibrium, the NCA has been used as an impurity solver for DCA \cite{Maier2000a,Maier2000b,Jarrell2001} even in the superconducting and pseudo-gap phase. As NCA is based on an expansion around $\Delta_\KK=0$, the results are expected to be most reliable in the Mott phase. This is confirmed in Fig.~\ref{fig1}c, which shows a comparison to continuous-time Quantum Monte Carlo results \cite{Gull2008} for the probability $p_0$ of finding the system in the ``plaquette singlet state'' (i.e., the ground state of the isolated cluster), which   
reflects  
the crossover from metallic to insulating behavior. In the following we focus on the insulating phase at $U=12$. As a measure for the short-range spin correlations, we compute the spin correlations in the $2\times 2$ cluster model,
\begin{align}
S_\delta(t) = \frac{1}{N_\delta}\sum_{i,j: |i-j|=\delta} \langle \hat S^z_i(t) \hat S^z_{j}(t) \rangle,
\end{align}
where $S^z_\QQ=\frac{1}{L_c} \sum_{\KK,\QQ,\sigma}e^{-i\QQ\bm R_i} \frac{\sigma}{2} c_{\KK\sigma}^\dagger c_{\KK+\QQ\sigma}$ is the spin operator at site $i$, and $N_\delta = \sum_{i,j: |i-j|=\delta} 1$ is the number of sites with distance $\delta$. 
The 
nearest and next nearest neighbor spin correlations,  $S_\text{NN} \equiv S_{\delta=1}$ and $S_\text{NNN} \equiv S_{\delta=2}$ clearly show a crossover to strong antiferromagnetic correlations below temperature $T\approx0.5$ (Fig.~\ref{fig1}b). At high temperature, the solution becomes similar to single-site DMFT \cite{Gull2008}, which is also evident in the spectral function (Fig.~\ref{fig1}a). With increasing temperature, the gap slightly shrinks, and the sub-structure of the Hubbard bands is washed out 
\cite{footnote01}.
One can also see directly  that the $\KK$-dependence of $\Sigma_\KK$ is lost with increasing $T$.

\begin{figure}[tbp]
\centerline{%
\includegraphics[width=\columnwidth]{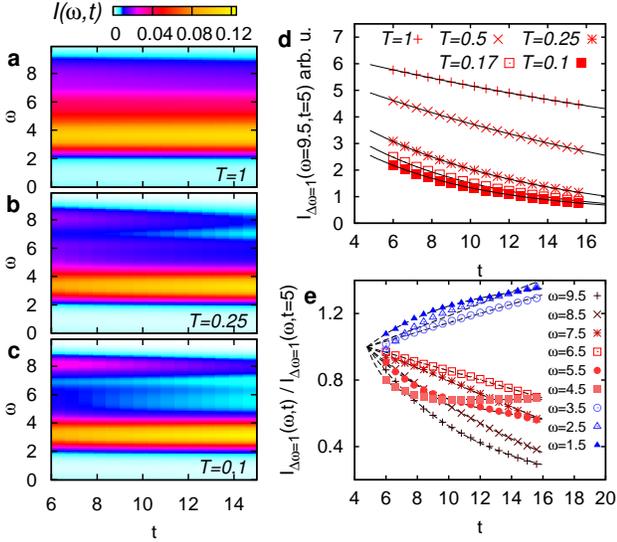}}
\caption{
a-c) Time- and frequency dependent photo-emission spectrum in the upper Hubbard band.
(Half-filled insulator at $U=12$, amplitude $E_0=6$, $N_\text{ex}\approx0.9\%$), for three different temperatures. 
d) PES weight integrated in the high-energy window $9\le \omega \le 10$ for various 
temperatures. The solid lines are exponential fits.
e) Frequency dependence of the relaxation for low temperatures, $T=0.1$.
These curves are normalized at $t=5$.
}
\label{fig2}
\end{figure}

{\bf Results --}  
We now simulate the time evolution of the insulating phase at $U=12$ after excitation with a laser pump of frequency $\Omega=16$, which populates states in the upper Hubbard band. Figures \ref{fig2}a-c show the time and energy-resolved photo-emission spectrum (PES)  $I(\omega,t)$ for three different temperatures. The PES is computed from the local (momentum-averaged) Green's function, using a Gaussian probe pulse with envelope 
$\propto \exp[-t^2/2\Delta t^2]$ 
and time resolution $\Delta t=3$, i.e., energy resolution $1/\Delta t$ \cite{FreericksKrishnamurthyPruschke2009}. The excitation density $N_\text{ex}$ can be defined by the integrated photoemission weight in the upper Hubbard band, where the total weight $\int \!d\omega \,I(\omega,t)$ is normalized to the particle density. For $U=12$ and the pump with $\Omega=16$ we find a roughly constant absorption coefficient $N_\text{ex}/E_0^2\approx 0.00024$ up to $N_\text{ex}\approx 3\%$. 
 During the timescale of the simulation, we observe a small increase of the total weight in the upper Hubbard band (up to $3\%$ of $N_\text{ex}$). This indicates carrier multiplication processes which can be important if the Hubbard gap is comparable to the bandwidth \cite{WernerAuger2014}. For the following discussion these effects are not dominant and will not be studied systematically. 
\begin{figure}[tbp]
\centerline{%
\includegraphics[width=\columnwidth]{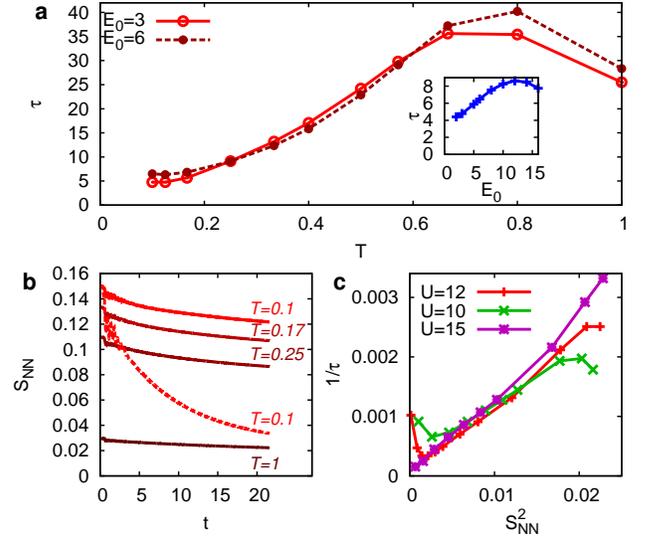}}
\caption{
a) Decay-time of high-energy PES at $U=12$ as a function of temperature at two fluences; the relaxation time is obtained from exponential fits shown in Fig.~\ref{fig2}d. Inset: Dependence of $\tau$ for temperature $T=0.1$ on the amplitude $E_0$. b) Time-dependence of $S_\text{NN}$ after excitation with low fluence ($E_0=6$, $N_\text{ex}\approx 0.9\%$ solid lines), and high fluence ($E_0=14$, $N_\text{ex}\approx 4.5\%$, dashed line). c) Relaxation rate $1/\tau$ in the low fluence regime ($E_0=3$, $N_\text{ex}\approx 0.2\%$) computed for various temperatures and plotted as a function of $S_\text{NN}^2$. For each value of $U$, relaxation times have beed obtained from the time-dependent weight $I_{\Delta\omega=1}(\omega,t)$ around the upper band edge ($U=10$: $\omega=8.5$,  pump frequency $\Omega=13$; $U=12$: $\omega=9.5$, $\Omega=16$; $U=15$: $\omega=10.75$, $\Omega=18$).
}
\label{fig3}
\end{figure}

We first focus on the low-fluence regime (excitation density $N_\text{ex}\lesssim1\%$), and discuss the fluence dependence later. On the timescale of the simulation, the redistribution of spectral weight from high to low frequency is clearly more pronounced at low temperature ($T=0.1$, Figure \ref{fig2}c) as compared to high temperature ($T=1$, Figure \ref{fig2}a). To analyze the redistribution of weight within the upper band, we integrate the PES over a given energy interval $\Delta \omega$, $I_{\Delta \omega}(\omega,t) = \int_{\omega-\Delta\omega/2}^{\omega+\Delta\omega/2} \!d\omega\, I(\omega,t)$. To understand the main role of various scattering channels we first focus on the decay of the high energy states, which shows an exponential time-dependence even on the short times simulated here (Fig.~\ref{fig2}d). The curves are fitted with a single-exponential decay $I(t) = A \exp(-t/\tau) + B$, and the relaxation rates are plotted in Fig.~\ref{fig3}a. The relaxation rate increases with temperature with a maximum at very high temperatures. 

As mentioned in the introduction, a possible candidate for relaxation is the scattering with short-ranged spin fluctuations. The disturbance of the short-range spin correlations and associated transfer of energy to the spin sector can be seen on a qualitative level through a decrease of the nearest-neighbor spin correlations $S_\text{NN}(t)$ (Fig.~\ref{fig3}b), which becomes more prominent with increasing fluence. For a carrier in a static antiferromagnetic background, it is possible 
to estimate a maximum energy transfer rate from mobile carriers to the spin sector \cite{Eckstein2014Layer}: For each hopping the average number of spin-flips against the antiferromagnetic exchange $J$ is proportional to the order parameter $m$, and hence the maximal spin-flip rate $\dot m/m$ is proportional to a hopping time. Consequently, the maximum rate of change of the spin energy $Jm^2$ is proportional to $m^2$. Motivated by this argument, we plot the temperature-dependent relaxation rate $1/\tau$ of the high energy part of the PES against $S_\text{NN}^2$ (Fig.~\ref{fig3}c). We observe a linear scaling in the intermediate temperature range, 
which leads to a main result of our work, 
namely 
that the inverse relaxation time can be taken as a measure of spin-correlations in the paramagnetic phase.  
The saturation and maximum at large temperatures (small $S_\text{NN}$) can be interpreted as a deviation from the simple spin picture. In agreement with this, we observe at these temperatures a slight filling in of the gap (see Fig.~\ref{fig1}a), and the saturation effect is shifted to larger temperatures as the interaction is increased from $U=10$ to $U=15$ (Fig.~\ref{fig3}c). 

\begin{figure}[tbp]
\centerline{%
\includegraphics[width=0.9\columnwidth]{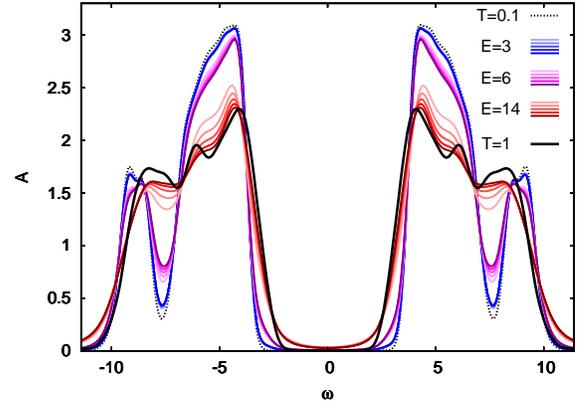}}
\caption{
The Figure shows the time-dependent spectral function $A(\omega,t)$ for the half-filled insulator at 
$U=12$, $\beta=10$ for various fluences. ($E_0=3$, $N_\text{ex}\approx 0.2\%$; 
$E_0=6$, $N_\text{ex}\approx 0.9\%$;
$E_0=14$, $N_\text{ex}\approx 4.5\%$), and various times
(t=$6,8.4,10.6,12,14.4$ from light to dark colors).
The thick black solid line is the spectral function at high temperature $T=1$, the 
dashed line  is the spectral function at he initial temperature $T=0.1$. For high fluences, 
the spectrum around the Hubbard bands becomes more similar to the high-temperature spectrum
at half-filling. However, the spectrum does not indicate thermalization. The larger number of mobile 
carriers of the photo-induced state is reflected in the filling of spectral weight in the gap.
}
\label{figsupp}
\end{figure}

While we have so far looked at the relaxation of the high-energy PES, it is worthwhile to briefly discuss how the measured relaxation time depends on the observable. Within the simulated time window the time-evolution of the weight $I(\omega,t)$ is not an exponential relaxation for all frequencies (Fig.~\ref{fig2}). At intermediate frequencies, e.g., weight is scattered both into and out of the states, leading to a non-monotonous time-evolution. For the same reason, also the time-evolution of averaged quantities such as the kinetic energy is not purely exponential, although the relaxation of these quantities also reflects the slow-down with increasing temperature. Another experimentally interesting feature is the dependence of the relaxation time on the excitation density. For large fluences we observe a rapid decrease of spin-correlations after the pulse (Fig.~\ref{fig3}b). Consistent with this observation and the dependence of the relaxation time on spin-correlations, we observe an increase of $\tau$ with fluence (Fig.~\ref{fig3}a, inset). (With the melting of short-range spin-fluctuations, also the spectral function around the Hubbard peaks becomes similar to the high-temperature spectral function, cf. Fig.~\ref{figsupp}).

\begin{figure}[t]
\centerline{%
\includegraphics[width=\columnwidth]{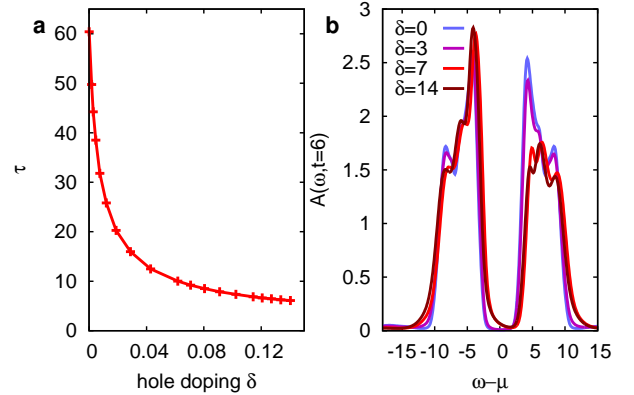}}
\caption{
a) Dependence of the relaxation-time on doping, for $U=12$, temperature $T=0.5$, and fluence $E_0=3$. The relaxation time is measured by an exponential fit to $I_{\Delta \omega=1}(\omega,t)$ in the high-energy regime $\omega-\mu=9.5$. b) Corresponding spectral functions.}
\label{fig4}
\end{figure}

Finally, we briefly discuss the effect of doping on the relaxation rates. At finite hole doping $\delta$ one would enter the pseudo-gap or superconducting regime at low temperature, where a simple approximation like NCA becomes less reliable. However, an important qualitative effect of doping can 
already 
be seen at high-temperature: Figure \ref{fig4} shows that the high-energy relaxation time rapidly decreases with doping, to a few inverse hoppings. This is opposite to the effect of spin-correlations, since spin-correlations generally decrease 
with increasing $\delta$. However, for finite $\delta$ additional mobile carriers provide an additional relaxation channel, in which a large number of chemically-doped carriers with low kinetic energy exchange energy with the small number of photo-doped carriers. 

In conclusion, we have used the nonequilibrium extension of DCA to study the effect of non-local spin correlations in the paramagnetic Mott phase of the Hubbard model on the relaxation of photo-doped carriers. Our results show that the relaxation times of high-energy photo-doped carriers, measured by the decay of the photo-emission signal close to the upper edge of the Hubbard band, reflect the short-range spin correlations in the paramagnetic phase. Taking parameters appropriate for cuprates, the relaxation times range from $\approx 10-20$fs for low temperatures to $\approx 200$fs when spin correlations are suppressed. In future studies, it will be interesting to resolve the momentum and time dependent scattering at lower energies (which requires slightly longer simulation times), to see how not only nearest-neighbor spin correlations, but more generally the full momentum and frequency-dependence of spin fluctuations may be obtained from the time-resolved photo\-emission (or two-photon photoemission) spectrum. More generally, the results show that short-range spin correlations act as a dissipative environment for mobile carriers, which may also be probed in cold atom experiments. 

\acknowledgments
We thank K. Balzer, J. Bon\v{c}a, C. Gianetti, D. Gole\v{z}, A.~Lichtenstein, Z.~Lenar\v{c}i\v{c}, A.~Millis, P. Prelov\'sek, N.~Tsuji
and L.~Vidmar  for fruitful discussions and E. Gull for providing the QMC data of Ref.~\cite{Gull2008}. 
ME acknowledges the Aspen Center for Physics and the NSF Grant No.~1066293 for 
hospitality. The calculations were run on the supercomputer HLRN of the North-German 
Supercomputing Alliance. PW acknowledges support from FP7/ERC starting grant No. 278023.


\begin{thebibliography}{45}
\bibitem{footnote02}
For an introduction to the Keldysh formalism, as well as concerning the notation for 
Keldysh Green's functions, self-energies and Dyson equations, the reader is referred to Ref.~\cite{REVIEW}.
\end{thebibliography}
\end{document}